\documentclass[prc,twocolumn,secnumroman,tightenlines,amssymb,aps,nobibnotes,
               superscriptaddress,balancelastpage,floatfix,nofootinbib]{revtex4}
\usepackage{graphicx}
\usepackage{multirow}           
\usepackage{color}	
\renewcommand{\labelitemi}{-}
\expandafter\ifx\csname package@font\endcsname\relax\else
\expandafter\expandafter
\expandafter\usepackage
\expandafter{\csname package@font\endcsname}
\fi
\DeclareRobustCommand\substyle{\name@idx{document substyle}}
\DeclareRobustCommand\classoption{\name@idx{document class option}}
\DeclareRobustCommand\classname{\name@idx{document class}}
\def\name@idx#1#2{{\ttfamily#2}
\index{#2\space#1=\string\ttt{#2}\space#1}\index{#1>#2=\string\ttt{#2}}}

\usepackage{enumitem}
\setitemize{itemsep=0pt}
\usepackage{hyperref}
\hypersetup{
    colorlinks=true, 
    linkcolor=blue,
    filecolor=blue,      
    urlcolor=blue,
    citecolor=blue,
    pdftitle={}
}
\begin{document}

\title{
$^{178}$Hg and asymmetric fission of neutron-deficient pre-actinides

}

{\author{A. Jhingan}
\affiliation{Inter University Accelerator Centre, Aruna Asaf Ali Marg, Post Box 10502, New Delhi 110067, India}
\author{C. Schmitt}
\email{christelle.schmitt@iphc.cnrs.fr}
\affiliation{Institut Pluridisciplinaire Hubert Curien, CNRS/IN2P3-UDS, 67037 Strasbourg Cedex 2, France}
\author{A. Lemasson}
\affiliation{GANIL, CEA/DRF-CNRS/IN2P3, BP 55027, 14076 Caen Cedex 5, France}
\author{S. Biswas}
\affiliation{GANIL, CEA/DRF-CNRS/IN2P3, BP 55027, 14076 Caen Cedex 5, France}
\author{Y.H. Kim}
\affiliation{Center for Exotic Nuclear Studies, Institute of Basic Science, 55 Expo-ro, Daejeon, 34126, Korea}
\author{D. Ramos}
\affiliation{GANIL, CEA/DRF-CNRS/IN2P3, BP 55027, 14076 Caen Cedex 5, France}
\author{A.N. Andreyev}
\affiliation{Department of Physics, University of York, York YO10 5DD, United Kingdom}
\affiliation{Advanced Science Research Center, Japan Atomic Energy Agency, Tokai, Ibaraki 319-1195 Japan}
\affiliation{ISOLDE, CERN, CH-1211 Geneve 23, Switzerland}
\author{D. Curien}
\affiliation{Institut Pluridisciplinaire Hubert Curien, CNRS/IN2P3-UDS, 67037 Strasbourg Cedex 2, France}
\author{M. Ciema{\l}a}
\affiliation{The Niewodniczanski Institute of Nuclear Physics - PAN, 31-342 Kraków, Poland}
\author{E. Clément}
\affiliation{GANIL, CEA/DRF-CNRS/IN2P3, BP 55027, 14076 Caen Cedex 5, France}
\author{O. Dorvaux}
\affiliation{Institut Pluridisciplinaire Hubert Curien, CNRS/IN2P3-UDS, 67037 Strasbourg Cedex 2, France}
\author{B. De Canditiis}
\affiliation{Institut Pluridisciplinaire Hubert Curien, CNRS/IN2P3-UDS, 67037 Strasbourg Cedex 2, France}
\author{F. Didierjean}
\affiliation{Institut Pluridisciplinaire Hubert Curien, CNRS/IN2P3-UDS, 67037 Strasbourg Cedex 2, France}
\author{G. Duchêne}
\affiliation{Institut Pluridisciplinaire Hubert Curien, CNRS/IN2P3-UDS, 67037 Strasbourg Cedex 2, France}
\author{J. Dudouet}
\affiliation{Univ. Lyon, UCLB 1, CNRS/IN2P3, IP2I Lyon, UMR 5822, F-69622, Villeurbanne, France}
\author{J. Frankland}
\affiliation{GANIL, CEA/DRF-CNRS/IN2P3, BP 55027, 14076 Caen Cedex 5, France}
\author{G. Frémont}
\affiliation{GANIL, CEA/DRF-CNRS/IN2P3, BP 55027, 14076 Caen Cedex 5, France}
\author{J. Goupil}
\affiliation{GANIL, CEA/DRF-CNRS/IN2P3, BP 55027, 14076 Caen Cedex 5, France}
\author{B. Jacquot}
\affiliation{GANIL, CEA/DRF-CNRS/IN2P3, BP 55027, 14076 Caen Cedex 5, France}
\author{C. Raison}
\affiliation{Department of Physics, University of York, York YO10 5DD, United Kingdom}
\author{D. Ralet}
\affiliation{GANIL, CEA/DRF-CNRS/IN2P3, BP 55027, 14076 Caen Cedex 5, France}
\author{B.-M. Retailleau}
\affiliation{GANIL, CEA/DRF-CNRS/IN2P3, BP 55027, 14076 Caen Cedex 5, France}
\author{L. Stuttgé}
\affiliation{Institut Pluridisciplinaire Hubert Curien, CNRS/IN2P3-UDS, 67037 Strasbourg Cedex 2, France}
\author{I. Tsekhanovich}
\affiliation{Univ. Bordeaux, CNRS, CENBG, UMR 5797, F-33170 Gradignan, France}
\author{A.V. Andreev}
\affiliation{Joint Institute for Nuclear Research, 141980 Dubna, Russia}
\author{S. Goriely}
\affiliation{Institut d’Astronomie et d’Astrophysique, ULB, CP 226, BE-1050 Brussels, Belgium}
\author{S. Hilaire}
\affiliation{CEA, DAM, DIF, F.-91297, Arpajon, France}
\author{J.-F. Lemaître}
\affiliation{CEA, DAM, DIF, F.-91297, Arpajon, France}
\author{P. Möller}
\affiliation{Mathematical Physics, Lund University, 221 00 Lund, Sweden}
\author{K.-H. Schmidt}
\affiliation{Rheinstraße 4, 64390 Erzhausen, Germany}

\begin{abstract}
\noindent
Fission at low excitation energy is an ideal playground to probe the impact of nuclear structure on nuclear dynamics. While the importance of structural effects in the nascent fragments is well-established in the (trans-)actinide region, the observation of asymmetric fission in several neutron-deficient pre-actinides can be explained by various mechanisms. To deepen our insight into that puzzle, an innovative approach based on inverse kinematics and an enhanced version of the VAMOS++ heavy-ion spectrometer was implemented at the GANIL facility, Caen. Fission of $^{178}$Hg was induced by fusion of $^{124}$Xe and $^{54}$Fe.  The two fragments were detected in coincidence using VAMOS++ supplemented with a new SEcond Detection arm. For the first time in the pre-actinide region, access to the pre-neutron mass and total kinetic energy distributions, and the simultaneous isotopic identification of one the fission fragment, was achieved. The present work describes the experimental approach, and discusses the pre-neutron observables in the context of an extended asymmetric-fission island located south-west of $^{208}$Pb. A comparison with different models is performed, demonstrating the importance of this ``new" asymmetric-fission island for elaborating on driving effects in fission. 
\end{abstract}

\maketitle

\section{Introduction}

Due to its relevance for fundamental physics, impact in astrophysics, and importance for a variety of technological and societal usage, fission at low excitation energy is an intense field of nuclear research since its discovery in the late 30's \cite{meitner:1939}, both on the experimental and theoretical front (see Refs.~\cite{andreyev:2018, schmidt:2018, bender:2020} for recent reviews). First focused on fissioning actinides for cross section and possible application reasons, these studies established that the nascent fragment shell structure is a crucial driving force in deciding the fission split. That permitted to go beyond the pioneering theory \cite{bohr:1939} based on a purely macroscopic ``liquid-drop"-like picture.

Fission involves obviously a complex re-arrangement of a many-body quantum system made of two types of nucleons. Due to the difficulty in precisely identifying the fission products, for several decades mostly fragment mass distributions with limited resoluton were available. Hence, the inferred respective roles of the proton and neutron sub-systems remained model-dependent to large extent with no firm experimental validation. Additional measurements of the total kinetic energy ($TKE$) and the number of neutrons ($M_n$) emitted by the fragments after scission were found critical to reach a deeper understanding of the process. Yet, their dependence on the concomitant influence of both fragments, and the role of specific ``magic" nucleon numbers could not be un-ambiguously resolved. In parallel, and combined with increase in computing resources, fundamental theories developed. However, approaches based on various, sometimes contradictory, assumptions could describe existing experimental data equally well, leaving many aspects un-settled. Further insight could be initiated only recently, thanks to important progress in experimental technology. State-of-the-art detection combined with the use of inverse kinematics \cite{schmidt:2000, caamano:2015, pellereau:2017} gave access to precise information on the fission fragment isotopic composition, while radioactive beam facilities \cite{schmidt:2000, andreyev:2010} extended the knowledge about low-energy fission properties to a wider domain of the nuclear chart. The availability of precise isotopic information over the full production\footnote{Isotopic identification of the fragment was achieved already in Ref.~\cite{boucheneb:1989}, but only for a part of the production. In addition, the $Z$ was not uniquely resolved.}, and of a large variety of fissioning systems, triggered exciting theoretical developments regarding neutron and proton sharing between the fragments at scission and the evidence of potentially new driving effects \cite{ichikawa:2012, panebianco:2012, warda:2012, andreev:2013, donnell:2014, ichikawa:2019, scamps:2020, prasad:2021, schmitt:2021, verriere:2021a, verriere:2021b, mahata:2021, bernard:2022}.

To refine models further, and explain most recent measurements, requires new types of observables and correlations, on one side, and widening the investigation to other, possibly more ``exotic", fissioning systems, on the other side (see {\it e.g.} Refs.~\cite{albertsson:2021, schmitt:2021b}). The exclusive asymmetric character of the fragment mass distribution of $^{180}$Hg at barrier excitation energy \cite{andreyev:2010}, and the following confirmation of an island of asymmetric fission in its neighbourhood \cite{andreyev:2018}, still presents many challenges to theory. A recent comprehensive analysis \cite{mahata:2021} showed that neutron-deficient pre-actinides are a key to clarify un-explained aspects exhibited by fissioning actinides, and reach a consistent, possibly universal, picture of fission over the nuclear chart.

Measuring the fragment mass informs about the degree of asymmetry of the split, which is intimately related to the potential-energy landscape of the fissioning system. The latter is expected to be governed to large extent by the quantum effects in the nascent fragments on the way towards scission. Since the potential energy has contributions from both fragments, and $A$ = $N + Z$, it is impossible to ascertain which, among the two partners, on one side, and, among the two nucleon sub-systems, on the other side, decides the mass partition. Measurements of $Z$ and $TKE$ allow a more selective investigation of the role of possibly specific proton-driven configurations. However, similar to the mass yield, these observables depend on the two fragments. Charge polarization, which is a measure of the neutron-richness of the fragments, and is customarily quantified by the deviation of the fragment charge $\Delta Z$ or $N/Z$ ratio from the unchanged-charge-density (UCD) assumption \cite{vandebosch:1973}, further helps in discriminating between the influence of the neutrons and protons. Yet, due to obvious conservation laws, neutron-richness of one of the fragments implies neutron-deficiency of its companion, preventing to separate the influence of the two partners. The number of neutrons $M_n$ evaporated by a fragment promptly after scission is given by its excitation energy. The latter is mainly contributed by the deformation energy at scission, which transforms into intrinsic excitation of the fragment along its shape relaxation to the ground state. It is therefore a signature of the influence of the fragment emitter. Though, neutron and proton effects both affect the fragment binding energy. Finally, since none of the available observables depends exclusively on a single nucleon sub-system of a specific fragment, unravelling un-ambiguously what drives fission requires the combinations of several observables. Such kind of complete data sets appeared recently for fission of actinides (see Refs.~\cite{ramos:2020, martin:2021} and references therein).

\begin{figure*}[!htb]
\includegraphics[width=0.95\linewidth]{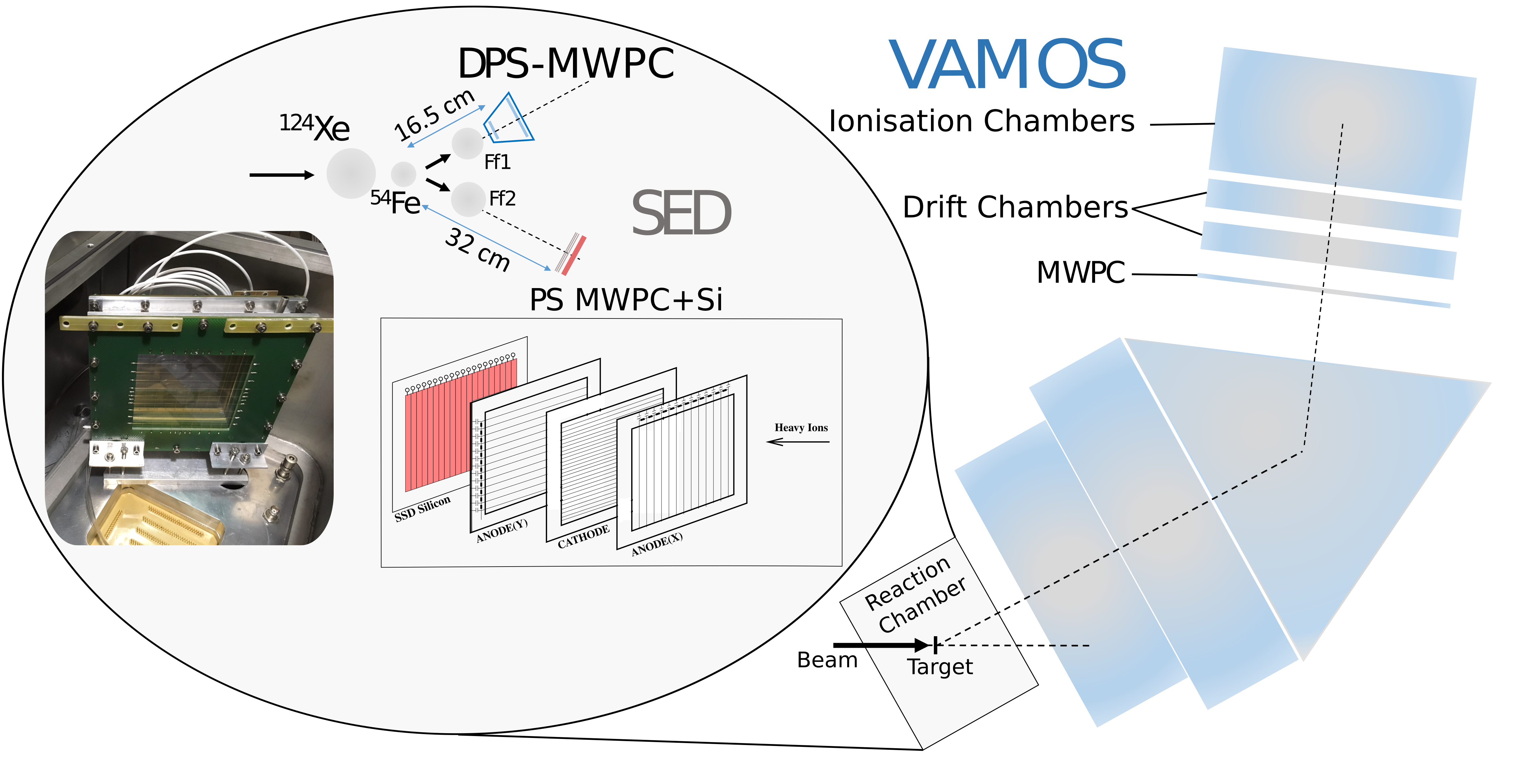}
\caption{Schematic of the experimental set-up for detecting the coincident fission partners (Ff1 and Ff2) at the enhanced VAMOS++ detection system. The general layout is shown on the right. A zoom of the target area with the VAMOS++ dual PS-MWPC and the new SED is given on the left. An exploded view of the SED and a photograph of its implementation in the reaction chamber are shown as well.}
\label{fig0}      
\end{figure*}

As a conclusion from the above, the critical need of {\it i)} accurate fragment identification, in both their neutron and proton contents, {\it ii)} simultaneous measurement of various observables, and {\it iii)} of a large variety of fissioning systems, are necessary to improve current understanding. In this context, the present work focuses on fission of $^{178}$Hg within an innovative approach implemented at the GANIL facility. Fission of pre-actinides as studied close to $\beta$-stability by Itkis {\it et al.} \cite{itkis:1990s} in the 90's, and in most recent works triggered by the observation of Ref.~\cite{andreyev:2010} on the neutron-deficient side, mainly consists in integral fragment-mass distributions with limited resolution. $TKE$ measurements were made available in several cases also. We refer to Ref.~\cite{mahata:2021} for an exhaustive list of the existing work, and to Refs.~\cite{bogachev:2021, kozulin:2022, dhuri:2022} published in the meantime. Scarce information on nuclear charge $Z$ exists \cite{boutoux:2013, gorbinet:2014}. In the present work, a unique data set  was collected by enhancing the VAMOS++ spectrometer of GANIL with a new SEcond Detection (SED) arm. The implementation of the latter was essential, providing the following three main 
advantages:
\renewcommand{\labelitemi}{$\circ$}
\begin{itemize}
\item clean selection of the events of interest as critical for the lowly fissile pre-actinide region;
\item determination of the integral pre-neutron fragment mass $A_{pre}$ and $TKE$ distributions together with the post-neutron mass $A_{post}$ and charge $Z$ from the heavy-ion spectrometer;
\item information of the $N/Z$ neutron-richness of the fragments at the moment of split and their post-scission neutron multiplicity $M_n$.
  \end{itemize}

The above  advantages will be described in details further in the text. The new set-up provides a large set of observables for a fissioning system located in a poorly explored region. In our previous letter \cite{schmitt:2021b}, the physics revealed by the new $N/Z$ and $M_n$ observables was highlighted, shedding further light into leading effects in fission across the nuclear chart. The present work communicates in detail about the experimental strategy and the specificities of the set-up, and discusses the ``standard" $A_{pre}$ and $TKE$ observables in the context of the asymmetric-fission island situated south-west of the well-established actinide island. A comparison with available models is also presented.

\section{Experimental approach}\label{M}

Accessing new observables with good precision as well as their correlations is particularly challenging for low-energy fission in the neutron-deficient lead area. A first difficulty is related to statistics, due to the low fission probability of pre-actinides. To partly circumvent this problem, fusion-induced fission has been shown in the last few years to be a good alternative to the ideal $\beta$-delayed and electromagnetic induced mechanisms, although the excitation energy of the fissioning system is somehow higher. While the width of the symmetric- and asymmetric-fission peaks varies with excitation energy, their position in the fragment mass (equivalently, charge) distribution does almost not, since shell effects are a property of a nucleus {\it per se}. In other words, their positions are expected to coincide for the pre-neutron distribution in $\beta$-delayed, electromagnetic, and fusion-induced fission. Still another challenge in the region is the requirement of new and higher-precision observables, which is made difficult by the relatively low kinetic energy of the fragments inherent to the fissioning system production mechanism. Finally, to trace back the situation at scission, which is the ``closest" one can approach the fission process, the coincident measurement of the two fragments is necessary.

\subsection{The enhanced VAMOS++ set-up}\label{Ma}

Fission of the neutron-deficient $^{178}$Hg nucleus was induced by fusion in inverse kinematics at GANIL. A $^{124}$Xe beam at 4.3 MeV/u impinged on a 130 $\mu$g/cm$^2$ thick $^{54}$Fe target evaporated on a $\approx$ 25 $\mu$g/cm$^2$ thick carbon backing, producing the $^{178}$Hg compound with an excitation energy $E^*$ = 34 MeV. A schematic layout of the set-up is given in Fig.~\ref{fig0}. The VAMOS++ magnetic spectrometer \cite{rejmund:2011}, placed at 29$^\circ$ with respect to the beam, was used to detect one of the fragments. The new SED arm \cite{akhil:2018} was installed 32 cm from the target on the other side of the spectrometer at an angle of 35$^\circ$ for the coincident measure of the fission partner. The angles of the two detection systems, and the central magnetic rigidity $B\rho$ of the spectrometer, which optimize efficiency and representativeness of the detected events, were determined based on reaction kinematics.

In front of the first quadrupole of VAMOS++, 16.5 cm away from the target, a dual position sensitive multi-wire proportional counter (PS-MWPC) \cite{vandebrouck:2016} gave access to the fragment emission angle as well as the start for the time-of-flight (ToF). About 760 cm downstream, following the magnetic elements, the 1 m-wide focal plane of the spectrometer was composed of a MWPC providing the stop signal of a first ToF, two drift chambers for $B\rho$ and trajectory reconstruction, and a segmented ionization chamber for energy loss and residual energy measurement. Further details on the VAMOS++ detection used in the present study are given in Ref.~\cite{ramos:2020}.

The new SED arm consisted of a two-dimensional PS-MWPC detector backed with a silicon strip detector (SSD), both of 10 $\times$ 10 cm$^2 $ active area. The PS-MWPC provided the stop of a second ToF (with respect to the VAMOS++ start) and the ($X$, $Y$) position of the coincident fission partner, while the SSD measured its energy. As compared to previous designs \cite{akhil:2021, akhil:2009}, a salient feature of the current SED system is the stacking of a transmission type low pressure PS-MWPC followed by a SSD in the same detector housing, {\it i.e.} a single aluminium chamber filled with isobutane at a pressure of 4 mbar. The PS-MWPC has a three-electrode geometry with the central timing cathode (for ToF) sandwiched between two position-sensitive anodes. Position information is extracted using the delay-line technique. The reduced wire pitch of 0.317 mm for the timing electrode and 0.635 mm for the position electrodes significantly improves the avalanche gains and timing resolutions \cite{akhil:2021}. Another salient feature is the integration of the timing pre-amplifiers with the detector body, eliminating cables between them. The position resolution of the PS-MWPC was found to be 1.2 mm (FWHM), while its intrinsic timing resolution has been estimated to be 200~ps~(FWHM)~\cite{akhil:2021}. The second layer consists of a 300 $\mu$m thick SSD (model TTT12 from Micron Semiconductors) with 20 strips on the front side (each 4.8 mm wide and 97 mm long) and interstrip separation of 50 $\mu$m. Readout is solely done on the back side, by means of a 24 pin FRC single inline connector. An energy resolution of 70 keV (FWHM) was observed for the 8.37 MeV $\alpha$ line of $^{230}$Th. A 0.9 $\mu$m Mylar foil is used as the entrance window for isolating the isobutane gas region of the SED from the high vacuum of the reaction chamber. The detector assembly was designed and prepared at the Inter University Accelerator Centre (IUAC), New Delhi, India, before shipment and implementation at GANIL.

The data acquisition was triggered by the coincidence between the entrance and focal plane timing signals of VAMOS++, the SED working as a slave. Calibration of times and energies was done using elastic scattering of, respectively, $^{54}$Fe in VAMOS++ and $^{124}$Xe in the SED, taking into account the appropriate energy loss on each line. The Si energy signal was additionally corrected for pulse height defect \cite{moulton:1978}.

\subsection{Data analysis}\label{Mb}

\begin{figure}[!htb]
\hspace{-1.cm}
\includegraphics[height=6.cm, width=1.05\columnwidth]{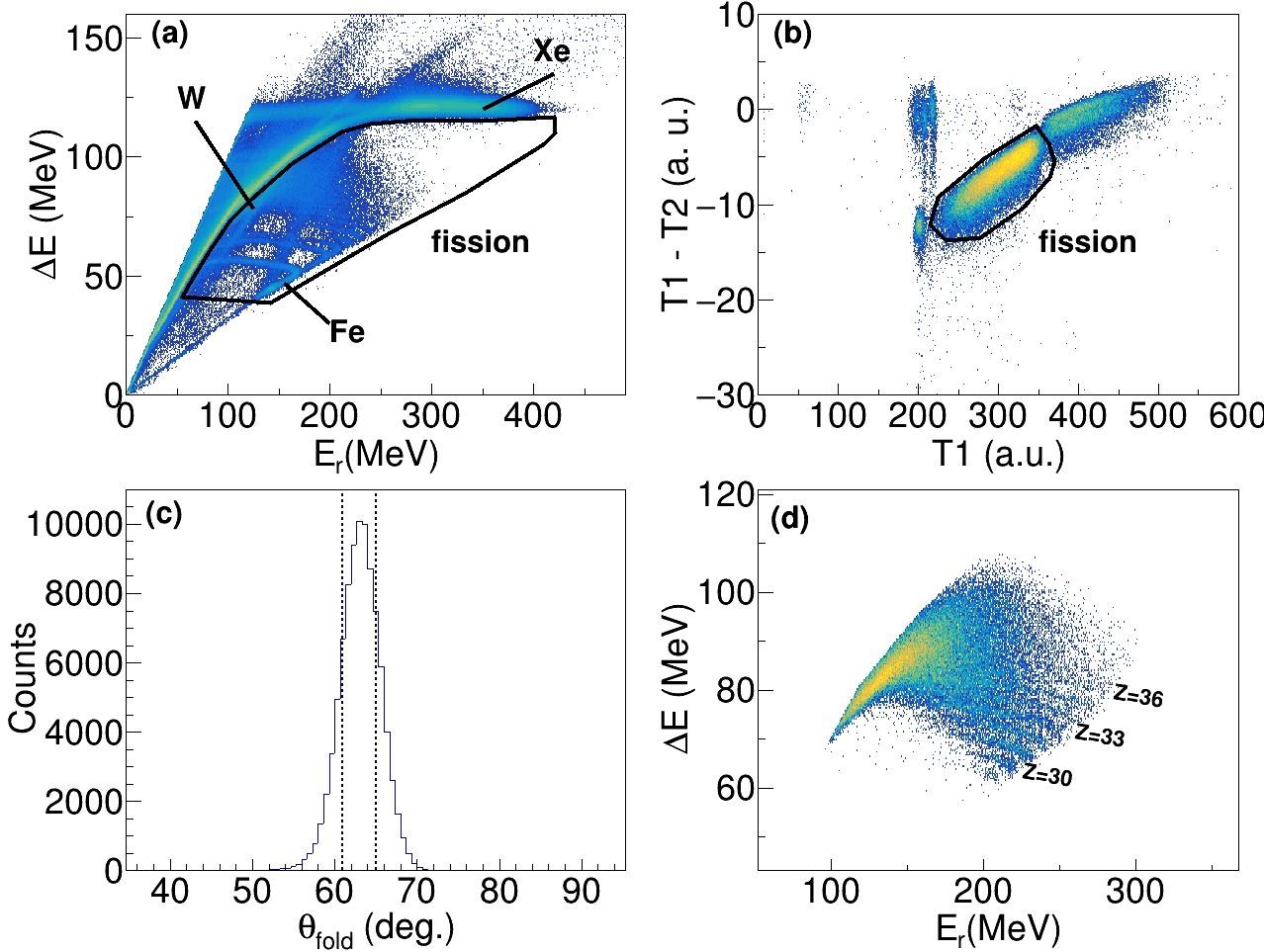}
\caption{(a) Correlation between the energy loss $\Delta E$ (in the first 3 segments of the ionisation chamber) and residual energy $E_r$ (in segments 2 to 6) measured at the focal plane of VAMOS++. The intense (green) lines are due elastic scattering events. The black contour delineates the region expected to be populated by the fission events of interest. (b) Correlation between the time-of-flight T1 of the fragment entering VAMOS++ and the difference in time-of-flight T1-T2 of the fragment detected by the SED and the fragment in VAMOS++, for those events satisfying the selection criterion of panel (a). The black contour delineates the area populated by fission. (c) Fission-fragment folding angle $\theta_{fold}$ distribution for those events satisfying the selection criteria of panels (a) and (b). Vertical lines delineate the peak due to fission. (d) ($\Delta E$, $E_r$) correlation for those events satisfying the selection criteria of panels (a), (b) and (c). Some $Z$ lines are indicated for reference.}
\label{fig1}      
\end{figure}

As compared to previous studies at VAMOS++ based on highly asymmetric beam-target combinations and fissioning actinides ({\it e.g.} Refs.~\cite{navin:2014, caamano:2015, ramos:2020}), the high probability of unwanted reactions and random coincidences complicates the selection of fission events in the present case. The dominance of background events is due to the more symmetric entrance channel and the lower fission cross section. Figure \ref{fig1}(a) displays the correlation between the energy loss $\Delta E$ and residual energy $E_{r}$ as given by the ionisation chamber for VAMOS++ singles. It is dominated by the intense lines, and associated tails, due to the elastic scattering off the $^{54}$Fe target and off tungsten impurities. The region expected to be populated by fission of $^{178}$Hg is delimited by the black contour. Contamination by a diffuse background is due to random events. Requiring the coincidence with the SED for events lying in this region leads to the spectrum of Fig.~\ref{fig1}(b) for the correlation between the times of flight of the ions detected on either side of the beam axis. Obviously, a substantial background mainly composed of remaining elastic events is still present: it appears as wings on both side of the fission region enclosed in the black contour in panel (b). The simultaneous application of the $\Delta E$-$E_{r}$ and ToF's gates provides a substantial reduction of contaminant events, as demonstrated with the plot of the fission-fragment folding angle $\theta_{fold}$ distribution which is centered around its expected mean value in panel (c). This observable permits still further cleaning up of the data set by setting a tight gate on the $\theta_{fold}$ peak, see vertical bars. Applying the three gates, {\it viz.} the $\Delta E$ {\it vs.} $E_r$, ToF's and $\theta_{fold}$ selection criteria, yields the ($\Delta E$, $E_{r}$) correlation displayed in panel (d). Compared to Fig.~\ref{fig1}(a), the efficient rejection of the unwanted reaction channels, and importantly of the diffuse background, is noteworthy. That demonstrates the first importance of the implementation of the new SED arm at VAMOS++ for the present physics case. Figure \ref{fig1}(d) contains a total of 6.8$\times 10^4$ coincidences, which are considered as true fission events and retained for further analysis. Note that the efficiency of the set-up amounts to about 2$\%$, given the kinematics of the reaction, the size of the detectors, and the spectrometer acceptance. 

The VAMOS++ spectrometer identifies with unique resolution the post-neutron mass $A_{post}$ ({\it i.e.} following cooling by evaporation after scission) and charge $Z$ of the fragment entering the spectrometer, and its velocity vector with high accuracy. The details of the analysis and performances of VAMOS++ for fission can be found in Refs.~\cite{rejmund:2011, kim:2017, ramos:2020}. Nuclear charge identification is obtained from the ($\Delta E$, $E_r$) correlation plot, where different $Z$'s populate distinct ``bands". Figure \ref{fig1}(d) shows that the latter can be well discriminated up to $Z = 38$. This is lower than the value reached in previous fission experiments at VAMOS++  ({\it e.g.} Refs.~\cite{navin:2014, caamano:2015, ramos:2020}), and is explained by the slower fragments produced in the reactions typically required to form neutron-deficient pre-actinides. Nuclear charge identification is very challenging for the involved nuclei having energies between 1 and 3 MeV/u, necessitating a compromise for the pressure of the ionisation chamber (20 mbar) to allow as heavy as possible elements not to end in the $Z$-unresolved Bragg region, on one hand, and to achieve good resolution for the lighter ones, on the other hand. Contrary to nuclear charge, the post-neutron mass identification is not impacted by the ($\Delta E$, $E_r$) limitation, as it relies essentially on the position and ToF measured on the VAMOS++ side \cite{navin:2014}. Very good resolution ($\Delta A_{post}$/$A_{post} \approx$ 0.8$ \%$) was achieved up to the heaviest fragments, including the Bragg region, as illustrated in Fig.~\ref{fig2}.

It is obvious from Fig.~\ref{fig2} that, even after integration over the whole  ($\Delta E$, $E_r$) matrix, the measured $A_{post}$ distribution is not symmetric about half the compound-nucleus mass as a physical $A_{post}$ spectrum should roughly be. This is due to the limited acceptance of the spectrometer, which is a complex function of emission angle and magnetic rigidity \cite{sugathan:2008}. For the present kinematics it is strongly related to the $Z$ of the fragment. An elaborate method was developed in Refs.~\cite{caamano:2015, ramos:2018} to correct for acceptance and recover the complete $A_{post}$ and $Z$ distribution yields. This method could not be applied to the present measurement in all its complexity due to limited statistics. However, for those events with $Z$ identified, the effect of acceptance can be accounted for with a simplified version of the most elaborate method.  Namely, we consider only those events within the same range in center-of-mass angle $\theta_{cm}$, over which the distribution is uniform. Under the assumption that fission is isotropic, this permits to recover the proper shape of the physical $A_{post}$ distribution for a given $Z$. Examples of isotopic distributions can be found in our earlier communication \cite{schmitt:2021b}. The selection on $\theta_{cm}$ implies a further reduction of the number of available events to 1.3$\times 10^4$. The results presented here below restrict to this subset, to ensure the absence of any bias due to acceptance effects.

\begin{figure}[!htb]
\includegraphics[height=6.cm, width=\columnwidth]{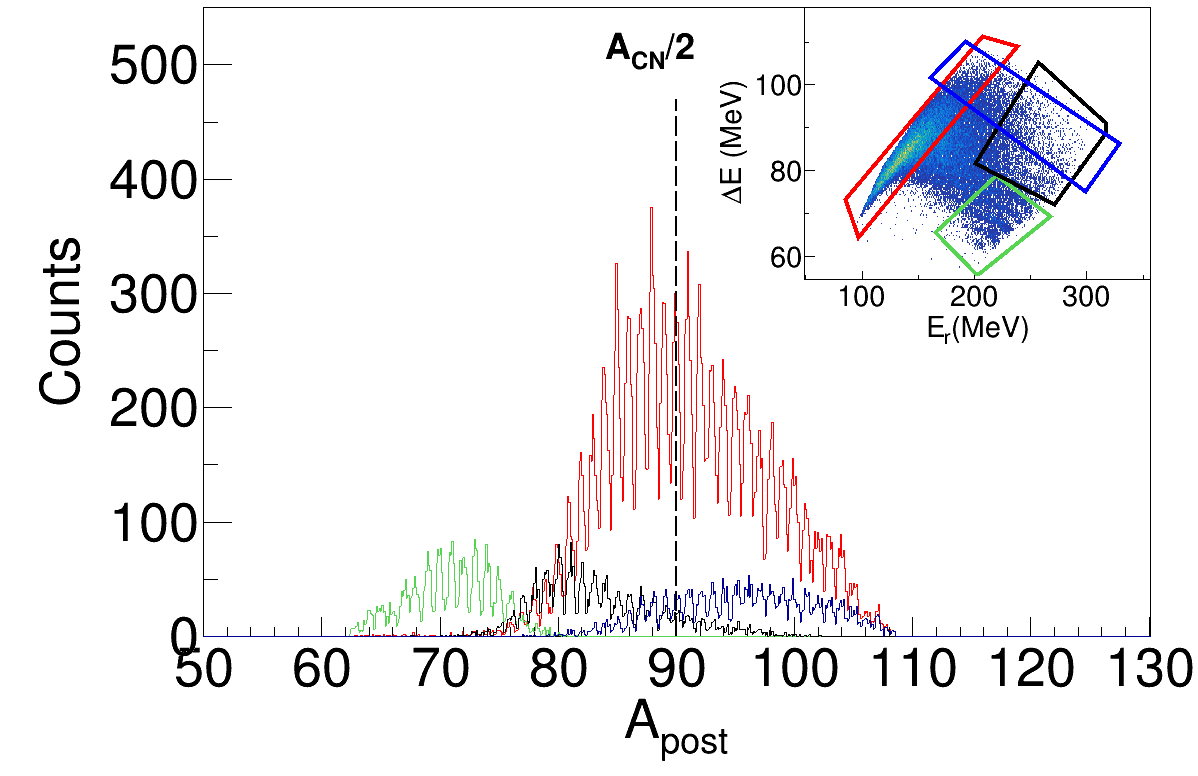}
\caption{Experimental post-neutron mass $A_{post}$ distribution for the fragments detected in VAMOS++. Different colors refer to ions populating different regions of the ($\Delta E$, $E_r$) correlation as defined in the inset. This matrix is identical to Fig.~\ref{fig1}(d).}
\label{fig2}      
\end{figure}

The velocity of the second fragment is derived from the timing and position signals provided by the PS-MWPC of the SED.  Combining the velocities of the coincident fragments, and assuming that evaporation by the compound nucleus before fission is negligible \footnote{Statistical model calculations were used to assess the reliability of this hypothesis \cite{schmitt:2021b}.}, the kinematical coincidence $2v$ method \cite{hinde:1992, choudhury:1999} can be applied to determine the pre-neutron mass $A_{pre}$, {\it viz.} of the fragments formed at the moment of scission, before de-excitation by neutron evaporation. The pre-neutron $TKE$ also follows from the measured velocities with $TKE$ = $0.5 A_{CN} v_{1 } v_2$ where $A_{CN}$ is the mass of the $^{178}$Hg compound nucleus, and $v_i$ is the velocity of fragment $i$ in the center of mass frame. The achieved resolution in pre-neutron mass and $TKE$ amounts to about 4 amu and 6 MeV (FWHM), respectively, primarily contributed by the short flight path on the second arm side. The enhancement of the set-up with the SED permits, to our knowledge, to apply the $2v$ method in fission for the first time with an advanced heavy-ion spectrometer such as VAMOS++. That demonstrates the second importance of the implementation of the new SED. Recently, a second arm was also installed at the heavy-ion PRISMA spectrometer for binary reaction studies \cite{galtarossa:2018}. The approach and data analysis,  which shares similarities with the present one, was so far applied to few-nucleon transfer channels in $^{197}$Au+$^{130}$Te collisions.

The innovative combination of the pre-neutron mass information with the isotopic yields was exploited to determine the neutron-richness $N$/$Z$ of the 
fragments at scission as well as the number of neutrons $M_n$ emitted per fragment promptly after scission. Till the present measurement, information about these signatures was non-existent for fissioning pre-actinides. Their availability is here due to installation of the second arm, demonstrating the third importance of the SED.

We note that the combination of VAMOS++ and the SED is in principle eligible to the $2v-2E$ method \cite{britt:1963, oed:1984}, and thus able to determine pre- and post-neutron masses after suitable corrections. Though, the present data analysis is based only on the $2v$ method as the $A_{post}$ capability of VAMOS++ overrides that of the SED. Combined with the $Z$ measurement provided by VAMOS++, it is the only way to extract the new $N$/$Z$ and $M_n$ observables with the required resolution.

\section{Results}\label{R}

The extraction of the $N$/$Z$ and $M_n$ observables and their significance were discussed in Ref.~\cite{schmitt:2021b}  to discriminate between the role of protons and neutrons, on one side, and specific scission configurations, on the other side. We focus here on the more ``standard" observables, {\it viz.} the pre-neutron $A_{pre}$ and $TKE$ distributions. These correspond to the bulk of information collected so far in the region for low-energy fission \cite{itkis:1990s, andreyev:2010, liberati:2013, ghys:2014, prasad:2015, tripathi:2015, nishio:2015, gupta:2019, gupta:2020, igort:2018, prasad:2021, gorbinet:2014, ghosh:2017, andel:2020, swinton:2021, nag:2021, bogachev:2021}. The present work supplements the existing set of systems with $^{178}$Hg, and discusses its features in the context of the asymmetric-fission island south-west of $^{208}$Pb. We note that $^{178}$Hg was investigated by Liberati {\it et al.} \cite{liberati:2013} in $\beta$-delayed fission. But only 8 events could be collected.

The experimental $A_{pre}$ and $TKE$ distributions, as well as their correlation are displayed in Fig.~\ref{fig4}. For the present Xe+Fe entrance channel, the question about a possible contribution from fast quasi-fission can be raised \cite{gupta:2020, kozulin:2021}. Within the acceptance of our set-up, and after application of the gates mentioned above, the selected events are confined around $\theta_{cm}$ = 81($\pm 2$)$^{\circ}$, thus minimizing the contamination, if any, by fast quasi-fission which is peaked forward and backward at near barrier energies (see {\it e.g.} \cite{hinde:2018} and references therein). Furthermore, according to the recent measurement by Bogachev {\it et al.} \cite{bogachev:2021} for similar reactions, wherever present, fast quasi-fission appears as distinct very asymmetric shoulders in the $A_{pre}$ distribution; such shoulders are not observed in the present measurement, see Fig.~\ref{fig4}(b). As far as slow quasi-fission events are concerned, since they imply a close to complete equilibration in mass and kinetic energy, {\it i.e.} approaching the compound nucleus configuration, their fission properties are expected to be close to those of fusion-fission \cite{khuyagbaatar:2015}, and thus not distort significantly the $A_{pre}$ and $TKE$ spectra. Finally, Time-Dependent-Hartree-Fock calculations \cite{simenel:2020} predict that quasi-fission is negligible for the present reaction. Consequently, we attribute the measured $A_{pre}$ and $TKE$ distributions as characteristic of fusion-induced fission of a $^{178}$Hg compound nucleus at $E^*$ = 34 MeV. 

\begin{figure}[!htb]
\includegraphics[height=5.cm, width=\columnwidth]{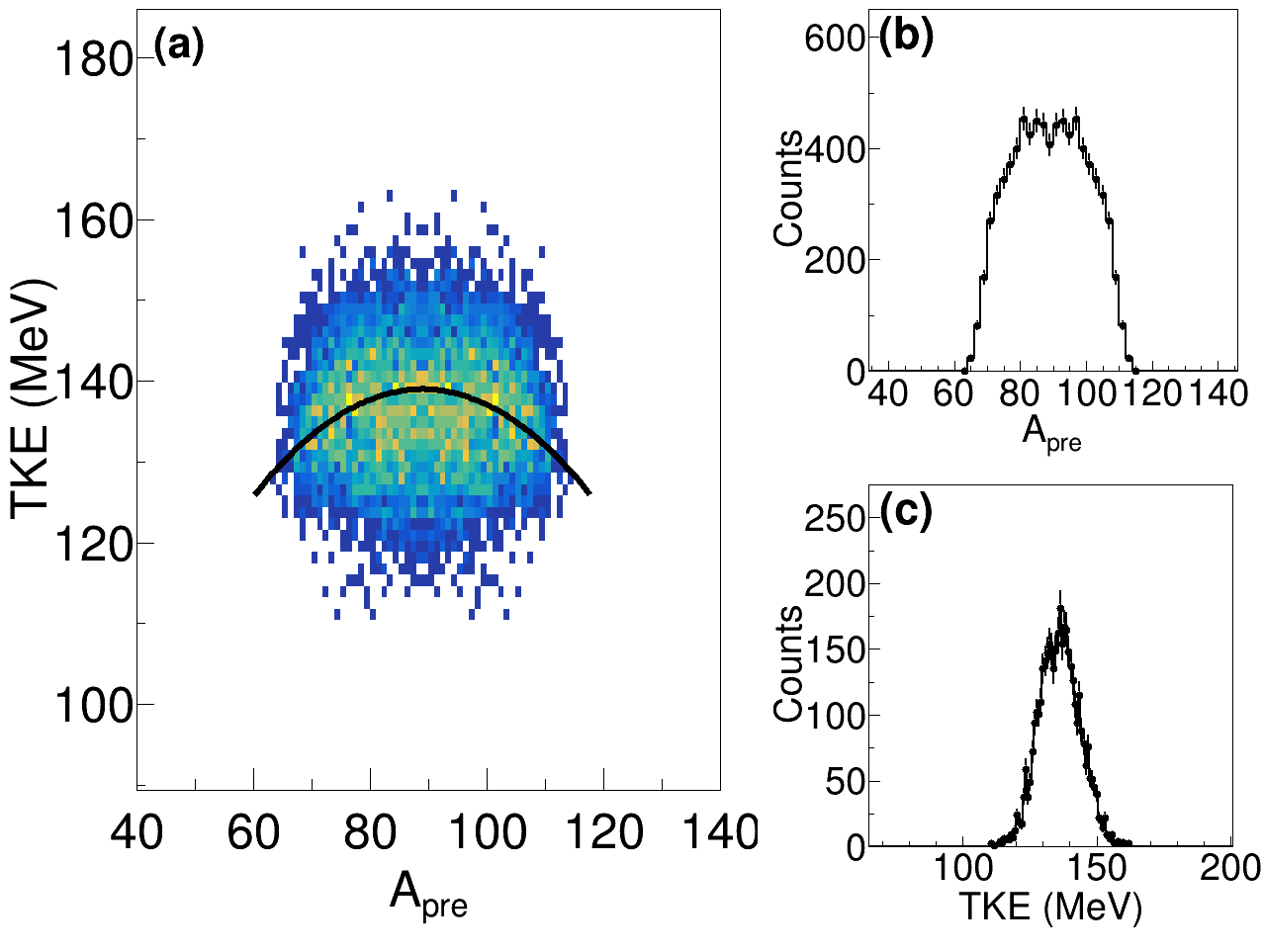}
\caption{Experimental ($A_{pre}$, $TKE$) matrix (a), and its projection on the $A_{pre}$ (b) and $TKE$ (c) axis. The solid line in (a) represents the Viola systematics~\cite{viola:1985}.}
\label{fig4}      
\end{figure}

The $A_{pre}$ distribution is seen to exhibit a broad shape with a flat top, and possibly a shallow dip at symmetry, suggesting the presence of both a symmetric and an asymmetric component. The pre-neutron $TKE$ distribution is single-humped, much resembling a Gaussian with mean value of 136 MeV and variance of 8 MeV, consistent with the compilation presented by Nishio {\it et al.} (see Fig.~3 of Ref.~\cite{nishio:2015}). 
We note that the $A_{pre}$ and $TKE$ distributions presented in this work slightly differ from those displayed in Ref.~\cite{schmitt:2021b}. This is due to a bit tighter gate on folding angle that was applied here to remove any remaining unwanted reactions. However, the main features are the same. Furthermore, the $N$/$Z$ and $M_n$ observables discussed in our earlier communication \cite{schmitt:2021b} are not affected by the small difference in $\theta_{fold}$ selection. The correlation between $A_{pre}$ and $TKE$ in Fig.~\ref{fig4}(a) is seen to exhibit the usual pattern, compatible with the Viola systematics \cite{viola:1985} extended to mass-asymmetric splits \cite{hinde:1987}. Unlike the observation reported for $^{178}$Pt \cite{igort:2018}, no elongated symmetric fission channel is evident at low $TKE$ in our data set, consistent with Prasad {\it {\it et al.}} \cite{prasad:2021}.

\section{Discussion}\label{D}

\subsection{Pre-actinide asymmetric-fission island}\label{Da}

The mass distribution obtained in this work for $^{178}$Hg (1.3$\times 10^4$ events) is compared in Fig.~\ref{fig5} to those of the close-by $^{180}$Hg and $^{178}$Pt systems investigated at similar excitation energy in Refs.~\cite{nishio:2015} and \cite{igort:2018}, respectively. Within the experimental error bars, the different mass resolutions, and the difference in the compound nucleus composition and excitation energy, the three data sets are observed to be very similar. Thus, $^{178}$Hg presents features essentially consistent with the so-far observed properties of asymmetric fission in the pre-actinide region. According to the similarity observed in Fig.~\ref{fig5} at intermediate $E^*$, and the dominantly asymmetric character of the mass spectrum of $^{180}$Hg at $E^*$ around the fission barrier \cite{andreyev:2010}, it is most likely that the asymmetric-fission component dominates too for $^{178}$Hg at low excitation, as was speculated from the 8 counts collected in Ref.~\cite{liberati:2013}. Based on so-far available empirical information, this conjecture suggests that $^{178}$Hg lies in the central part of the asymmetric-fission island whose boundary to the west, is thus still to be determined \cite{mahata:2021}.  

\begin{figure}[!ht]
  \includegraphics[width=\columnwidth]{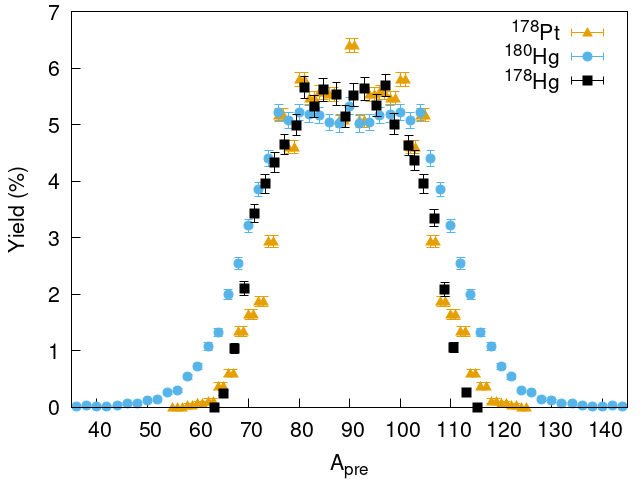}
  \caption{Experimental $A_{pre}$ distribution for $^{178}$Hg from this work (black squares), $^{180}$Hg from Ref.~\cite{nishio:2015} (light blue dots), and $^{178}$Pt from Ref.~\cite{igort:2018} (orange triangles). Error bars are of statistical nature. Experimental counts were normalized to 100$\%$.}
  \label{fig5}      
\end{figure}

To extract the contribution(s) of asymmetric fission, and the corresponding mean fragments masses, and investigate whether they coincide with stabilized nucleon configurations, it is customary to perform a multi-Gaussian fit analysis of the integral $A_{pre}$ distribution (see {\it e.g.} the aforementioned Refs.~\cite{nishio:2015, igort:2018}). Such an analysis is not done in this work, as we consider that it may not yield a unique solution. The latter can depend on a multitude of input aspects like the experimental conditions and data processing (resolution, target thickness, accuracy of energy loss corrections, among others), as well as the fitting procedure (number and choice of the free parameters, simultaneous adjustment of the $TKE$, etc). The ambiguity in the multi-Gaussian fit analysis regarding the amount and characteristic of possible competing fission channels or {\it modes} is illustrated by the conclusions drawn for the same fissioning system in different experiments. For example, while Nishio {\it et al.} \cite{nishio:2015} could explain the integral distribution of $^{180}$Hg at $E^* \approx$~33~MeV assuming only a single asymmetric mode occurs, Bogachev {\it et al.} \cite{bogachev:2021} concluded to the presence of up to three asymmetric modes in addition to a symmetric one. Similar discrepancies can be found for $^{182}$Hg and $^{178}$Pt in Ref.~\cite{kozulin:2022} and Ref.~\cite{prasad:2015}, and Ref.~\cite{igort:2018}, respectively. Very recently, Berriman {\it {\it et al.}} \cite{berriman:2022} discussed quantitatively the uncertainty of multi-Gaussian fits for a fissioning actinide.

\begin{figure}[!ht]
\includegraphics[width=\columnwidth]{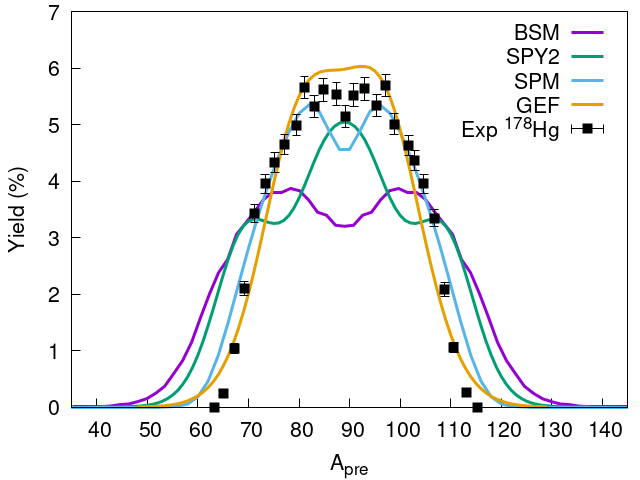}
\caption{Comparison between the experimental $A_{pre}$ distribution for $^{178}$Hg from this work (black dots) and various model calculations: BSM (violet), SPY2 (green), SPM (light blue), and GEF (orange). See the text for details. Experimental counts were normalized to 100$\%$.}
\label{figselection}      
\end{figure}
According to the possible uncertainty of the multi-Gaussian adjustment for limited-resolution experiments, we consider here that it is best suited to address the question of driving effects by discussing our measurement in connection with theory directly in terms of the {\it integral} $A_{pre}$ distribution. That is meaningful also since some models do not necessarily relate the measured distribution to specific fission valleys, but rather suggest an intricate competition between static effects and nuclear dynamics \cite{ichikawa:2012}.

\subsection{Comparison with theory}\label{Db}

The experimental $A_{pre}$ distribution is compared in Fig.~\ref{figselection} with four different calculations: 
the dynamical Brownian Shape Motion (BSM) model \cite{randrup:2011, moller:2012}, the microscopic scission point model (SPY2) \cite{lemaitre:2019, lemaitre:2021}, the improved macro-microscopic scission point model (SPM) \cite{andreev:2013}, and the semi-empirical GEneral Fission (GEF) model \cite{schmidt:2016, schmidt:2018} Version 2021/1.1. The theoretical curves were folded with the experimental resolution. However this was found to have no  effect on the comparison because the four calculations differ among each other by an amount which exceeds the experimental resolution.

The BSM and SPM models essentially predict the presence of asymmetric fission, with no distinct symmetric component, see Ref.~\cite{moller:2012} and Ref.~\cite{andreev:2013}, respectively. The five-dimensional potential energy landscape onto which the dynamical evolution of the fissioning nucleus is computed in BSM has no asymmetric valley but a deep symmetric channel (which is actually a fusion valley), see Fig.~7 of Ref.~\cite{ichikawa:2012}. The asymmetry in the calculated mass yield occurs when the nucleus, slightly beyond the second asymmetric saddle, slides down the side of a hill towards symmetry, but splits before reaching the bottom of the fusion
valley. It is therefore  un-related to shell structure expressed as a persistent valley extending from saddle to scission, which is a common feature in the calculated potential-energy surfaces of typical actinides \cite{ichikawa:2012}. The BSM-predicted yield curve is somewhat more asymmetric than seen in the experimental data. In the current implementation of the theory, the probability of changing the asymmetry when moving along a trajectory is independent of the neck diameter, which may lead to excessive asymmetry in cases like the present one. A similar behaviour was observed for $^{180}$Hg \cite{andreyev:2010}.

The position of the asymmetric fragment masses from SPM is consistent with experimental observation. In the improved scission-point model, the shell structure in the nascent fragments of the dinuclear scission configuration mostly defines the shape of the mass distribution. Due to Coulomb repulsion, the fragments at the scission point are strongly deformed. At symmetry, both fragments are close to the double-magic $^{90}$Zr, but the corresponding shell correction is positive. The considerable softness of the potential energy landscape in the fragment-deformation space can yield comparable corrections for asymmetric fragmentations. Combinations like $^{80}$Kr+$^{98}$Ru,  $^{82}$Kr+$^{96}$Ru, $^{82}$Sr+$^{92}$Mo, or $^{84}$Sr+$^{94}$Mo are particularly favored. We note that the account of the zero-point vibration energy also enhances the asymmetry of the mass distribution.

For both BSM and SPM, the yield at symmetry mainly originates from the filling of the dip between the asymmetric light and heavy peaks when their width gets broader with increasing excitation energy.

The SPY2 and GEF models expect a competition between symmetric and asymmetric fission, see Ref.~\cite{lemaitre:2019} and Ref.~\cite{schmidt:2016}, respectively. SPY2 predicts the dominance of symmetric over asymmetric splits, and the latter looks too asymmetric, similarly to BSM. The influence of fragments around $^{108}$Cd with 60 neutrons is mainly responsible for this partitioning. Though, with increasing mass of the fissioning nucleus, symmetric fission prevails. This is illustrated in Fig.~\ref{figspy2} which displays the fragment mass yield distributions for mercury isotopes as obtained with SPY2. Interestingly, $^{178}$Hg is located in the critical region of the transition between dominantly asymmetric and dominantly symmetric fission, and which is most sensitive to the influence of the excitation energy. Hence, it is a good test case to benchmark the model, and in particular the scission point distance \cite{lemaitre:2021}.

Based on an empirical analysis of the data available in 2014, the GEF code implements that fragments with proton number $Z$ around 36 play an important role in deciding the asymmetry of low-energy fission in neutron-deficient pre-actinides. Adoption in the model of a stabilizing effect around that $Z$ value also improved the description of actinide fission \cite{schmidt:2018}. Its existence was corroborated by several experiments since then, as well as by a recent extended systematics analysis and microscopic calculations, see Ref.~\cite{mahata:2021} and references therein. As reported in Fig.~5 of Gupta {\it et al.} \cite{gupta:2019}, GEF anticipates asymmetric fission to dominate at low $E^*$, and progressively weakens with increasing excitation at the benefit of an increase of symmetric splitting. Finally, a nearly flat top is reached around $E^*$ = 30-40 MeV. The GEF calculation is seen to describe reasonably well the experiment, and in particular the location of the asymmetric component in mass, similarly to SPM.

\begin{figure}[!ht]
\includegraphics[width=\columnwidth]{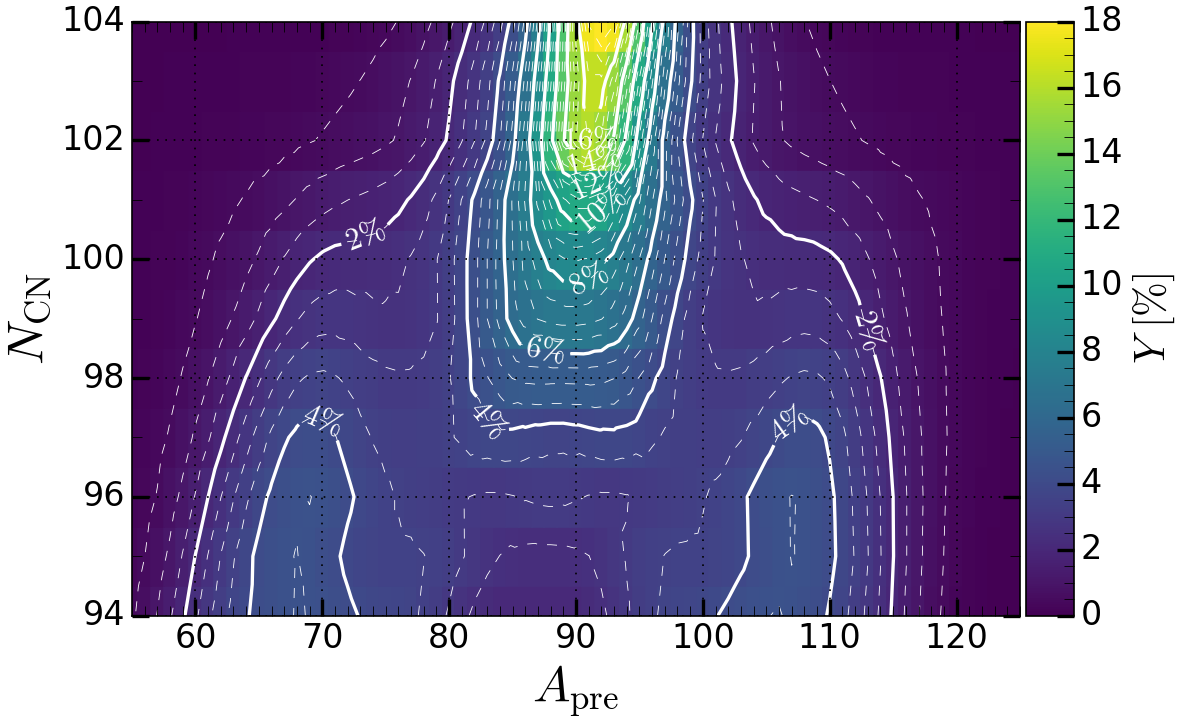}
\caption{Pre-neutron fission mass yield distributions for mercury isotopes with mass 174 to 184 at 34 MeV of excitation energy within the SPY2 model \cite{lemaitre:2021}.}
\label{figspy2}      
\end{figure}

According to the variety of assumptions and the uncertainty of some parameters, the ``inclusive" character of the $A$ and $TKE$ observables ({\it i.e.} none of the two depends on the $N$ or $Z$ of a specific fragment), and the fact that different models offer a reasonable gross description of their distributions, more exclusive observables are necessary to un-ambiguously figure out what drives asymmetric fission. Our recent letter \cite{schmitt:2021b} demonstrated that the $N$/$Z$ and $M_n$ quantities are particularly relevant in this respect. Unfortunately, predictions of these observables are not available today, due to the absence of experimental information, on one side, and the related theoretical difficulty, on the other side. This difficulty was challenged recently for a couple of fissioning actinides only, with the calculation of the $N$/$Z$ \cite{schmitt:2021, verriere:2021a, verriere:2021b} and $M_n$ \cite{albertsson:2021} observables. 

\section{Summary and conclusions}\label{Concl}

To study the interplay between structural and dynamical effects in low-energy fission in the challenging pre-actinide region, an innovative approach was implemented at GANIL, Caen. Fission of $^{178}$Hg was induced by fusion in inverse kinematics and an enhanced version of the VAMOS++ heavy-ion spectrometer was set up based on its coupling with a new SEcond Detection arm. It was used to detect in coincidence the two fragment products, determine the pre-neutron mass and $TKE$ distributions, the accurate isotopic identification of one of the partner and the number of neutrons emitted per fragment as a function of its charge. The new second arm was essential to {\it i)} select properly the fission events of interest out of the dominant background of unwanted reactions in the lowly-fissile region under discussion, {\it ii)} apply the kinematic coincidence $2v$ method in combination with a high-resolution mass and charge spectrometer, and thus {\it iii)} derive information on prompt neutron emission after scission. Such a data set is the first of this type for fission of a pre-actinide.

The present work focuses on the experimental approach, {\it viz.} the enhancement of VAMOS++ with the implementation of the SED, and on the discussion of the pre-neutron fragment-mass and $TKE$ observables. The mass distribution exhibits features of a mixed contribution of asymmetric and symmetric fission for $^{178}$Hg at an excitation energy of 34 MeV. Within the so-far existing systematics, the work suggests asymmetric fission to strongly dominate around the barrier for this nucleus. Thereby, it further expands the asymmetric-fission island in the pre-actinide region of the nuclear chart, leaving its left boundary still to be determined.

Comparison between experiment and different models  shows the relevance of studying pre-actinides for discriminating between different model approaches pertaining to driving effects in fission and their dependence on excitation energy. Though, according to the complex re-arrangement of the many-body neutron and proton sub-systems taking place in fission, the sole pre-neutron $A_{pre}$ and $TKE$ observables are not sufficient to draw an un-ambiguous conclusion. Experiments, going beyond conventional set-ups, are necessary. The present approach is a step in this direction, with the new arm enhancing the capabilities of the state-of-the-art heavy-ion spectrometer for the field. It is anticipated to be essential for unravelling the intricacies of the fission process.

\section{Acknowledgements}
We acknowledge the excellent support of the GANIL staff. We are grateful to M. Rejmund for help at various stages of the work. K. Nishio is thanked for providing us with the fusion-induced fission data of $^{180}$Hg, and C. Simenel is acknowledged for performing the entrance-channel TDHF calculations for our reaction. We appreciated fruitful discussions with G.N. Knyazheva. The work was partly sponsored by the French-German collaboration No. 04-48 between IN2P3/CNRS-DSM/CEA and GSI, the French-Polish agreements LEA COPIGAL (Project No. 5) and IN2P3-COPIN (Project No. 12-145), the LIA France-India agreement, and by STFC (UK).

\end{document}